\documentclass{article}
\usepackage{spconf,amsmath,graphicx}
\usepackage{mathrsfs}
\usepackage{makecell}
\usepackage{multirow}
\usepackage{booktabs}
\usepackage{subfigure}
\usepackage{bm}

\title{Synchronous Transformers for End-to-End Speech Recognition}
\name{Zhengkun Tian$^{1,2}$, Jiangyan Yi$^{1}$, Ye Bai$^{1,2}$, Jianhua Tao$^{1,2,3}$, Shuai Zhang$^{1,2}$, Zhengqi Wen$^{1}$}
\address{$^1$National Laboratory of Pattern Recognition, Institute of Automation, \\ Chinese Academy of Sciences, Beijing, China \\
$^2$School of Artificial Intelligence, University of Chinese Academy of Sciences, Beijing, China\\
$^3$CAS Center for Excellence in Brain Science and Intelligence Technology, Beijing, China}
\begin{document}
%
\maketitle
\begin{abstract}
For most of the attention-based sequence-to-sequence models, the decoder predicts the output sequence conditioned on the entire input sequence processed by the encoder. The asynchronous problem between the encoding and decoding makes these models difficult to be applied for online speech recognition. In this paper, we propose a model named synchronous transformer to address this problem, which can predict the output sequence chunk by chunk. Once a fixed-length chunk of the input sequence is processed by the encoder, the decoder begins to predict symbols immediately. During training, a forward-backward algorithm is introduced to optimize all the possible alignment paths. Our model is evaluated on a Mandarin dataset AISHELL-1. The experiments show that the synchronous transformer is able to perform encoding and decoding synchronously, and achieves a character error rate of 8.91\% on the test set.
 
\end{abstract}
\section{Introduction}
\label{sec:intro}

Attention-based sequence-to-sequence models \cite{bahdanau2014neural,vaswani2017attention,vinyals2015show,chorowski2015attention,kim2017joint,dong2018speech}, especially transformer model \cite{vaswani2017attention}, have shown great success in various tasks, e.g. neural machine translation \cite{bahdanau2014neural, vaswani2017attention}, image captioning \cite{vinyals2015show} and speech recognition \cite{chorowski2015attention,kim2017joint,dong2018speech}. 



For conventional attention-based sequence-to-sequence models, the inference process can be divided into two stages. The encoder first processes an entire input sequence into a high-level state sequence. After that, the decoder predicts the output sequence conditioned on the previous predicted symbol and context vector extracted from the entire encoded state sequence. This makes the models encode and decode sequences asynchronously, and prevents it from being applied for online speech recognition. There are some works trying to solve this problem. Tjandra et al. \cite{tjandra2017local} propose a local monotonic attention mechanism that forces the model to predict a central position at every decoding step and calculate soft attention weights only around the central position. However, it's difficult to accurately predict the next central position just based on limited information. Monotonic chunkwise attention \cite{chiu2017monotonic} is proposed to adaptively split the encoded state sequence into small chunks based on the predicted selection probabilities. But complex and tricky training methods make it hard to implement. Triggered attention  \cite{moritz2019triggered} utilizes the spikes produced by connectionist temporal classification (CTC) model to split the sequence into many state chunks, and then the decoder predicts the output sequence in a chunkwise way. However, triggered attention requires forced alignment to assist model training. Most of the proposed models introduce additional components and have very tricky training methods.

\vspace{-1pt}
In this paper, we propose a synchronous transformer model (Sync-Transformer), which can perform encoding and decoding at the same time. The Sync-Transformer combines the transformer \cite{dong2018speech} and self-attention transducer (SA-T) \cite{zhengkun2019sat} in great depth. Similar to the original transformer, the Sync-Transformer has an encoder and a decoder. In order to eliminate the dependencies of self-attention mechanism on the future information, we first force every node in the encoder to only focus on its left contexts and ignore its right contexts completely. Once a fixed-length chunk of state sequence is produced by the encoder, the decoder begins to predict symbols immediately. Similar to the Neural Transducer \cite{jaitly2015neural, jaitly2016online, sainath2018improving}, the decoder generates the output sequence chunk by chunk. However, restricted by the time-dependent property of RNNs, the Neural Transducer model only optimizes the approximate best alignment path corresponding to the chunk sequence. By contrast, we adopt a forward-backward algorithm to optimize all possible alignment paths and calculate the negative log loss function as the same as the RNN-Transducer \cite{graves2013speech} and SA-T \cite{zhengkun2019sat}. We evaluate our Sync-Transformer on a Mandarin dataset AISHELL-1. The experiments show that the Sync-Transformer is able to encode and decode sequences synchronously and achieves a comparable performance with the transformer.


\vspace{-1pt}
The remainder of this paper is organized as follows. Section 2 describes our proposed Sync-Transformer. Section 3 presents our experimental setup and results. The conclusions and future work will be given in Section 4.

\begin{figure*}[t]
	\centering
	\subfigure[The Structure of Synchronous Transformer and Inference Process]{
		\centering
		\label{fig:enc_and_infer}
		\includegraphics[width=0.45\linewidth]{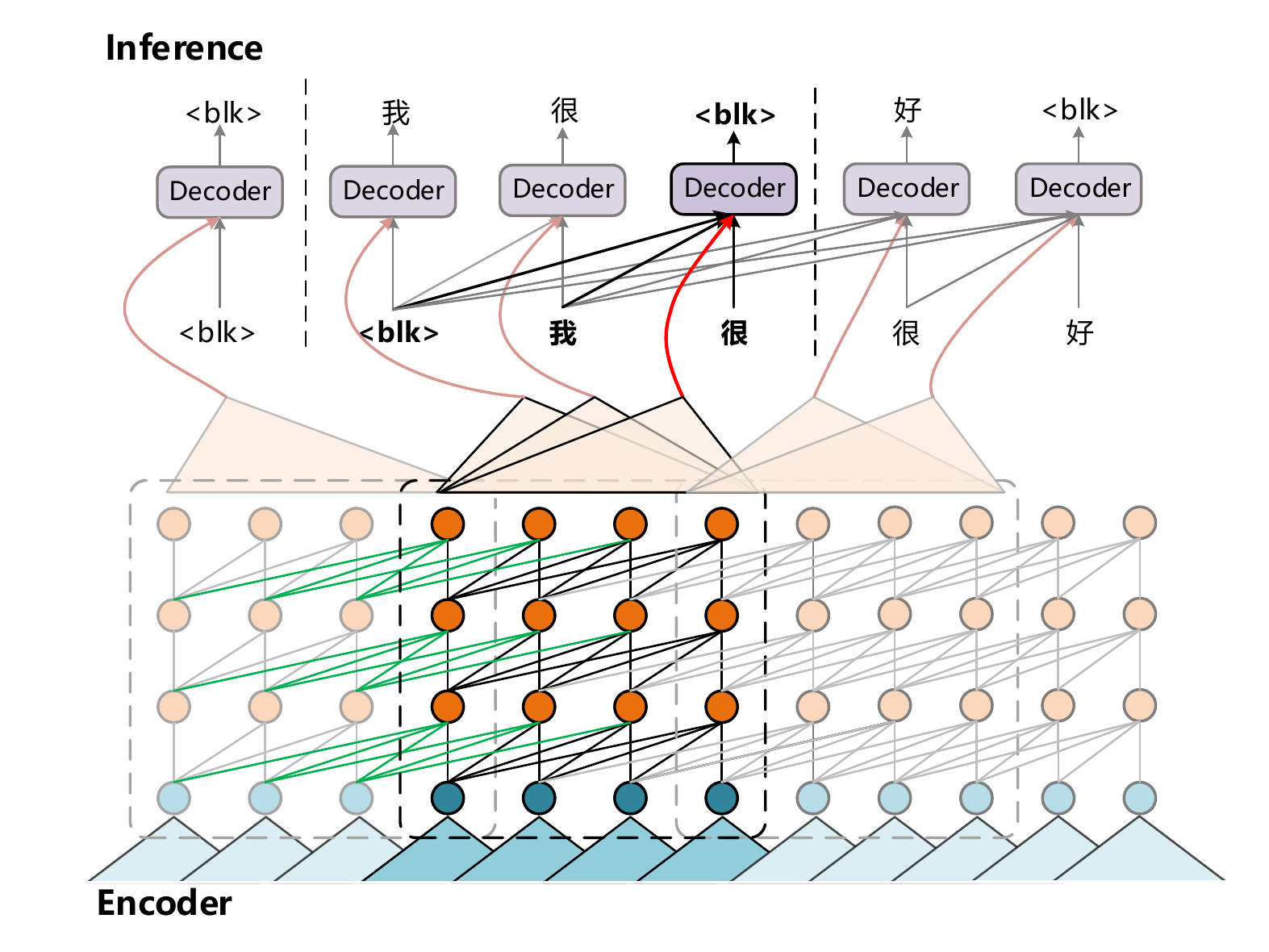}}
	\subfigure[The Structure of Decoder]{
		\centering
		\label{fig:decoder}
		\includegraphics[width=0.25\linewidth]{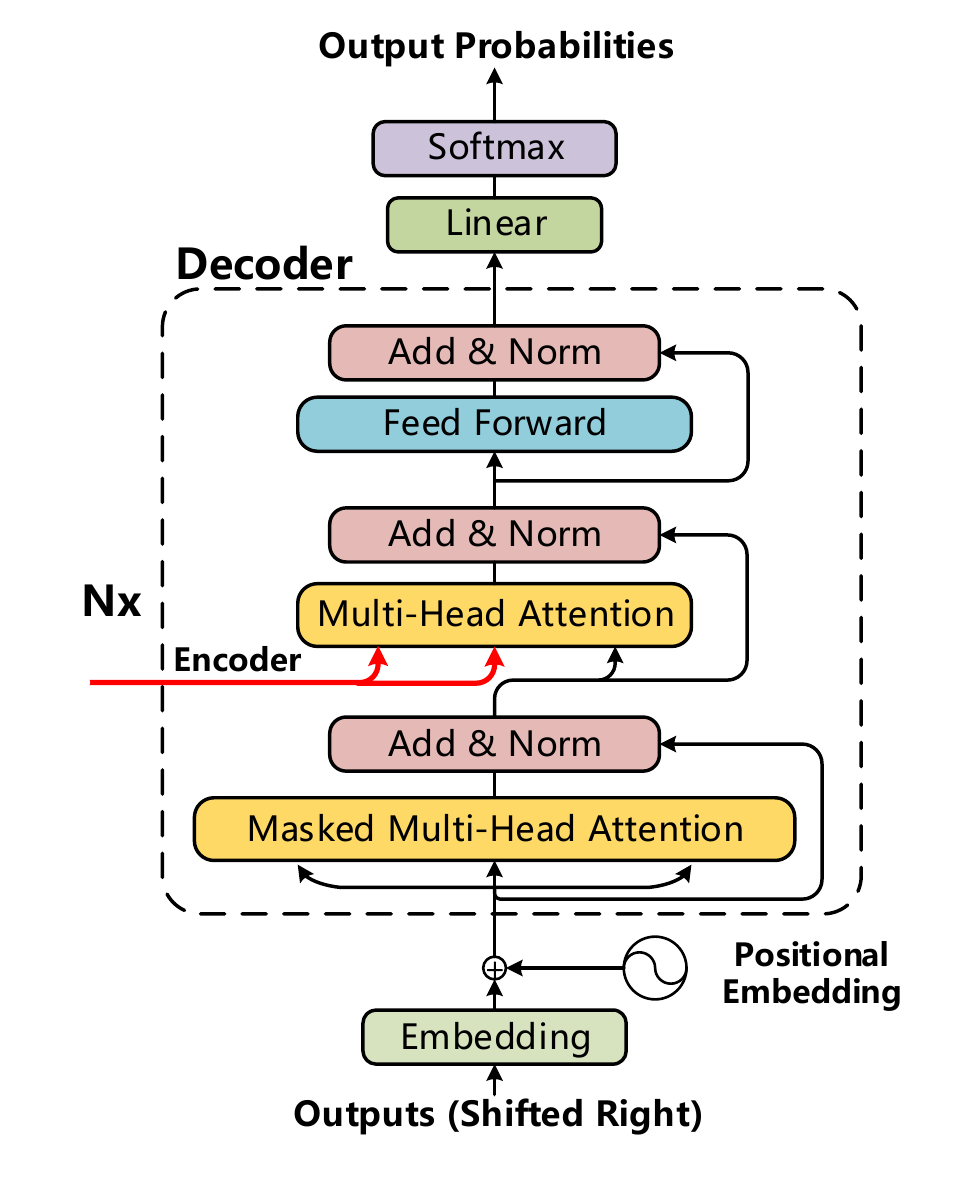}}
	\subfigure[Output Probability Lattice]{
		\centering
		\label{fig:latice}
		\includegraphics[width=0.2\linewidth]{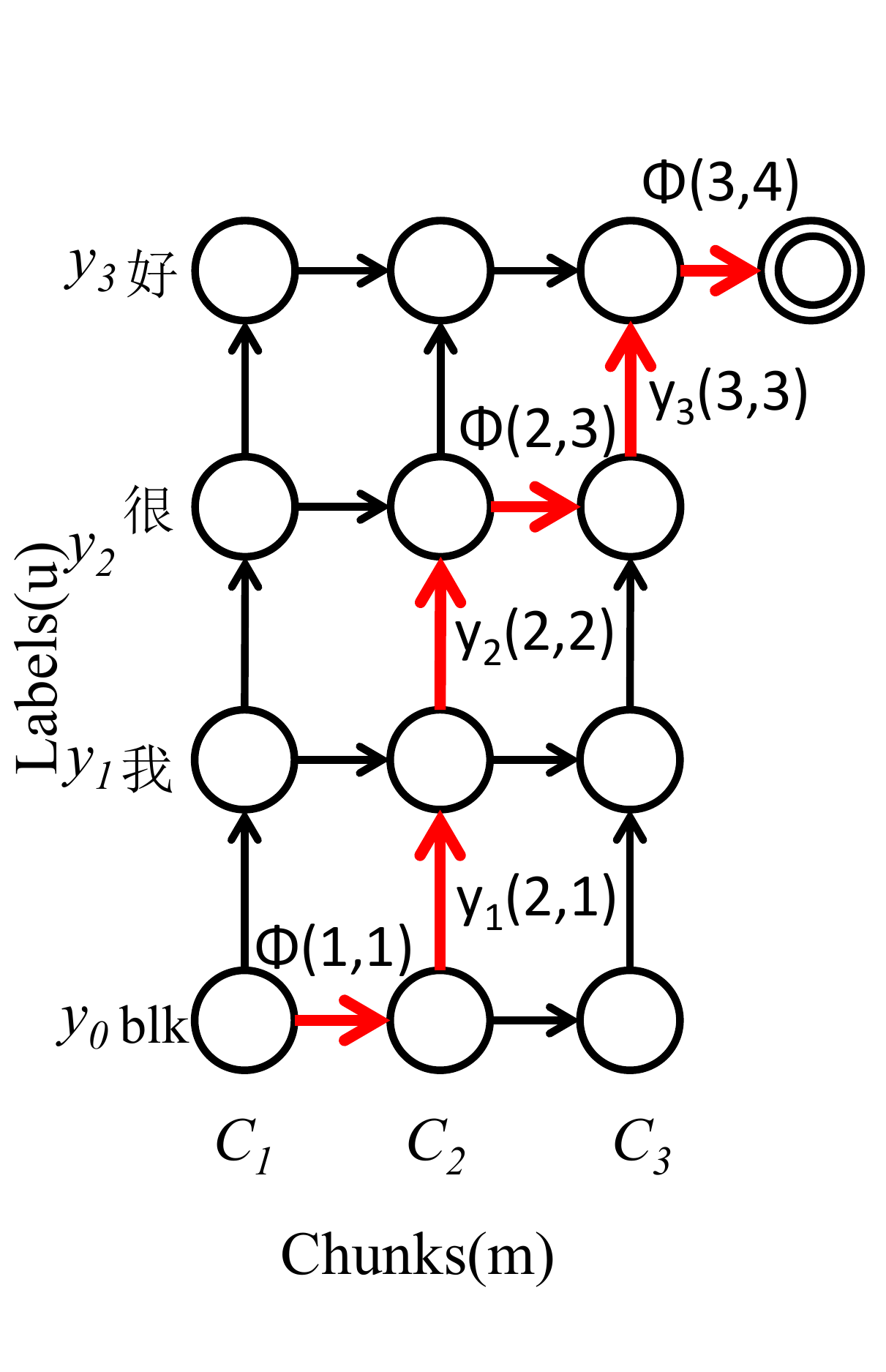}}
	\vspace{-5pt}
	\caption{(a) illustrates the whole structure of Sync-Transformer and the inference process. The Sync-Transformer consists of an encoder and a decoder. Every node in the encoder only pays attention to its left contexts. The decoder generates symbols chunk by chunk. Once a fixed-length chunk of sequence is processed by the encoder, the decoder begins to predict the output symbols immediately. (b) illustrates the details of the decoder. (c) illustrates a output probability lattice utilized to sum over the probabilities of all possible alignment paths by forward-backward algorithm. }
	\vspace{-10pt}
\end{figure*}

\section{Synchronous Transformer}

\subsection{Model}
Similar to the transformer \cite{vaswani2017attention}, a Sync-Transformer consists of an encoder and a decoder, as depicted in Fig.1(a). Both encoder and decoder are composed of multi-head attentions and feed-forward layers\cite{vaswani2017attention, dong2018speech, zhengkun2019sat}. 

As shown in Fig.1(a), we put a 2D convolution front end at the bottom of the encoder to process the input speech feature sequences simply, including dimension transformation(transform feature dimensions from 40 to 256), time-axis down-sampling and adding sine-cosine positional information \cite{vaswani2017attention}. Let $x_{1:T}$ be the input feature sequence, the processed sequence can be expressed as $s_{1:L}$, where $T$ and $L$ are the lengths of these two sequences respectively. 


In order to get rid of the dependencies on the entire input state sequences, we make the following modifications on the original self-attention encoder, as depicted in Fig.1(a). On the one hand, we force every node in the encoder to focus on its left context and ignore its right contexts completely during calculating self-attention weights. Although each intermediate node can only model local dependencies information, the top node of the encoder can still model long-term dependencies. On the other hand, similar to transformer-xl \cite{dai2019transformer}, the encoder of Sync-Transformer processes the input sequence chunk by chunk. There is an overlap between two adjacent chunks to maintain a smooth transition of information between chunks. For the processed input sequence $s_{1:L}$, the encoder can split it into $M$ encoded state chunks $C_{1:M}$. This means that the calculation of attention weights just depend on a $W$-length chunk rather than the entire input sequence. Let $B$ be the overlapping length of two adjacent chunks. The relationship between $M$ and $L$ can be expressed as $M=\lceil \frac{L-W}{W-B} + 1 \rceil $.
\begin{equation}
C_{1:M} = Encoder(s_{1:L})
\end{equation}

At every decoding step, the decoder predicts a symbol conditioned on the previous predicted symbols $y_{0:u-1}$ ($ 0 \leq u \leq U+1$) and one chunk. Once a $\langle \textit{blk} \rangle$ symbol is predicted, the decoder will switch to the next chunk and continue decoding. This process can be expressed by the following formula.

\begin{equation}
p(y_{u}|y_{0:u-1}, C_m)=Decoder(y_{0:u-1}, C_m)
\end{equation} 
where $C_m$ ($ 1 \leq m \leq M$) represents the $m$-th chunk produced by the encoder.
\vspace{-3pt}
\subsection{Training}
The training process is divided into two steps. In order to accelerate the convergence, we first use a trained transformer model to initialize the parameters of Sync-Transformer. Then apply the following forward-backward algorithm to train a Sync-Transformer.

The encoder splits the input sequence into $M$ chunks. The decoding process in every chunk ends with $\langle\textit{blk}\rangle$. It is difficult to figure out which chunk each target symbol should belong to. Therefore, we construct an output probability lattice graph as shown in Fig.1(c) by calculating the probabilities of all target symbols in each chunk. Given the input sequence $x_{1:T}$, the probability of the output $y_{1:U}$ is calculated by summing over the probabilities of all possible alignment paths.

\begin{equation}
p(y_{1:U}|x_{1,T}) =\sum_{y\in\mathcal{Y}}p(y_1,y_2,...,y_{(U+M)}|x_{1:T})
\end{equation}
Where $\mathcal{Y}$ represents the set of all possible alignment paths. It's intractable and inefficient to calculate the probability $p(y_{1:U}|x_{1,T})$ by enumerating all possible alignment paths. Therefore, like transducer-based models in \cite{graves2012sequence, zhengkun2019sat}, we introduce a forward-backward algorithm to calculate the probabilities efficiently.

The \textit{forward variable} $\alpha(m,u)$ means the sum of the probability of all the possible paths, which begin with the start symbol  $y_0$ ( = $\langle \textit{blk} \rangle$) and end with $y_u$ in $m$-th chunk. Given $m$-th chunk and the predicted symbol sequence $y_{0:u-1}$, the probabilities of predicting $\langle \textit{blk} \rangle$ and $y_{u}$ are represented as $\phi(m, u-1)$ and $y_{u}(m, u-1)$  respectively. For all $1 \leq m \leq M$ and $ 1 \leq u \leq U+1$, the forward variables can be calculated recursively using 


\begin{equation}
\begin{aligned}
\alpha(m,u)  &= \alpha(m-1,u)\phi(m, u) \\
&+ \alpha(m,u-1)y_{u}(m, u-1)
\end{aligned}
\end{equation}
And all paths begin with $y_0$ ( = $\langle \textit{blk} \rangle$), it means $\alpha(1,1)=1$. The probability $p(y_{1:U}|x_{1,T})$ can be expressed by the forward variable at the terminal node.
\begin{equation}
p(y_{1:U}|x_{1,T}) = \alpha(M,U+1)\phi(M, U+1)
\end{equation}

Similarly, the \textit{backward variable} $\beta(m,u)$ means the sum of the probabilities of all possible paths, which begin with $y_u$ in $m$-th chunk and end with $y_{U+1}$(=$\langle \textit{blk} \rangle$) in the last chunk. The backward variables can be expressed as 

\begin{equation}
\begin{aligned}
\beta(m,u)  &= \beta(m+1,u)\phi(m, u) \\
&+ \beta(m,u+1)y_{u+1}(m, u)
\end{aligned}
\end{equation}
where the initial condition $\beta(M,U+1)=\phi(M, U+1)$.

Given an input feature sequence $x_{1,T}$ and a target sequence $y_{1:U}$, the probability $p(y_{1:U}|x_{1,T})$ is equal to the sum of $\alpha(m,u)\beta(m,u)$ over any top-left to bottom-right diagonal through the nodes. That is, $\forall n: 2 \leq n \leq U+M+1$

\begin{equation}
p(y_{1:U}|x_{1,T})=\sum_{(m, u):m+u=n}\alpha(m,u)\beta(m,u)
\end{equation}

We train the model to minimize the negative log-loss function $\mathcal{L}=-\text{ln}p(y_{1:U}|x_{1,T})$. The calculation of gradients is exactly the same as RNN-T \cite{graves2012sequence}.

\subsection{Inference}
The inference process is displayed in Fig.1(a). During inference, the decoder will predict the output symbols conditioned on a fixed-length chunk of encoded state sequences and all the previous predicted non-blank symbols. it might predict one or more symbols in a chunk. Once a $\langle\textit{blk}\rangle$ is predicted, It will switch to the next chunk and continue decoding. The decoder will repeat the above steps till all the chunks are processed. To simplify the inference process, we don't try to merge some alignment paths with the same prefix.

\section{Experiments and Results}
\label{sec:pagestyle}

\subsection{Dataset}
\vspace{-3pt}
In this work, all experiments are conducted on a public Mandarin speech corpus AISHELL-1\footnote{http://www.openslr.org/13/} \cite{bu2017aishell}. The training set contains about 150 hours of speech (120,098 utterances) recorded by 340 speakers. The development set contains about 20 hours (14,326 utterances) recorded by 40 speakers. And about 10 hours (7,176 utterances) of speech is used to be test set. The speakers of different sets are not overlapped.

\vspace{-5pt}
\subsection{Experiment Setup}
\vspace{-3pt}

For all experiments, we use 40-dimensional Mel-filter bank coefficients (Fbank) features computed on a 25ms window with a 10ms shift. Each feature is re-scaled to have zero mean and unit variance for each speaker. We chose 4232 characters (including a blank symbol '$\langle\textit{blk}\rangle$' and a unknown symbol '$\langle\textit{unk}\rangle$' ) as model units.

We utilize Kaldi\footnote{https://github.com/kaldi-asr/kaldi} for data preparation. And our Sync-Transformer is built on ESPNet \cite{watanabe2018espnet} and warp-rnnt\footnote{https://github.com/1ytic/warp-rnnt}. It consists of 6 encoder blocks and 6 decoder blocks. There are 8 heads in multi-head attention. The 2D convolution front end utilizes two-layer time-axis CNN with ReLU activation, stride size 2, channels 256 and kernel size 3. The output size of the multi-head attention and the feed-forward layers are 256.  In order to accelerate the convergence, we replace the ReLU activation function in the feed-forward network with gated linear units \cite{dauphin2017language}. We empirically set the left context of every node in the encoder to 20 and the right context to 0. More context parameter settings will be explored in the future. What's more, we adopt an Adam optimizer with warmup steps 25000 and the learning rate scheduler reported in \cite{vaswani2017attention}.

During decoding, we use a beam search with a width of 5 for all the experiments. And set the maximum length of symbols generated in a chunk is 10. We use character error rate (CER) to evaluate the performance of different models.

\vspace{-3pt}
\subsection{Results}
\subsubsection{Comparison of different window lengths}
We first explore how chunks with different lengths $W$ affect the performance of Sync-Transformer. For every experiment, the overlapping range of adjacent chunks is set to 20\% of the chunk length. As shown in Table 1, the Sync-Transformer with chunk length 10 can achieve a CER of 9.06\% on the test set. When the length is reduced to 5, Sync-Transformer still performs well. However, if the length is greater than 20, it leads to severe performance degradation. We suppose that there may be more than one character in a long chunk, which might make the model difficult to predict the output sequence accurately.  

The length of speech segment represented by a fixed-length is $W\times4\times10ms$, where 4 means that 2D convolution front end can reduce the speech length by 4 times and 10ms represents the frame shift. The Sync-Transformer can achieve a competitive performance depending on a latency of 0.4s. If the overlap is taken into account, it is actually 0.32s.

When the chunk length is 1, Sync-Transformer is similar to a transducer model, which decodes a sequence frame by frame. In turn, when the length is large enough, there is one chunk for any utterances. In this case, it is equivalent to a transformer model.

\vspace{-5pt}
\subsubsection{Comparison of different overlap lengths}
\vspace{-5pt}
Next, we try to figure out the effects of overlap between chunks on the performance. Based on previous experiments, we set the length of the chunks to 10 for all experiments in this section. From Table 2, we find that the overlapping between the chunks plays an important role. The Sync-Transformer with overlap length 3 can achieve a CER of 8.91\% on the test set. When the overlap is set to 1 or 0, too little overlap between two adjacent chunks may be the main cause of the degradation of performance. Therefore, we suppose that a decent overlap can maintain smooth transition of information flow between the chunks. The performance of the model also decreases when the overlap is set to 4. We guess that the large overlap will cause the information contained in the two adjacent chunks to be very similar, which will further degrade the performance of the model.

\begin{table}[t]
	\caption{Comparison of different window lengths (CER \%).}
	\label{tab:table1}
	\centering
	\begin{tabular}{c|ccccc}
		\toprule
		$\bm{W}$ & \textbf{5} & \textbf{10} & \textbf{15} & \textbf{20} & \textbf{25} \\
		\hline
		\textbf{Dev} & 8.64 & \textbf{7.99} & 8.57 & 8.68 & 11.04 \\
		\textbf{Test} & 9.73 & \textbf{9.06} & 9.51 & 9.76 & 11.71 \\
		\bottomrule
	\end{tabular}
	\vspace{-15pt}
\end{table}

\begin{table}[t]
	\caption{Comparison of different overlap lengths (CER \%).}
	\label{tab:table2}
	\centering
	\begin{tabular}{c|ccccc}
		\toprule
		$\bm{B}$ & \textbf{4} & \textbf{3} & \textbf{2} & \textbf{1} & \textbf{0} \\
		\hline
		\textbf{Dev} & 8.60 & \textbf{7.91} & 7.99 & 9.53 & 9.61 \\
		\textbf{Test} & 9.56 & \textbf{8.91} & 9.06 & 10.39 & 10.47 \\
		\bottomrule
	\end{tabular}
	\vspace{-15pt}
\end{table}


\vspace{-5pt}
\subsubsection{Comparison with other end-to-end models}
\vspace{-5pt}
We also compare the Sync-Transformer with other end-to-end models. The transformer model is trained according to the recipe in ESPnet\cite{watanabe2018espnet}, which has the same settings as our Sync-Transformer. The second column indicates whether the model can decode in a streaming way. And the third column indicates the number of steps required to decode a $U$-length sentence. And $M$ is the number of chunks and $T$ is the number of speech frames.

The experiments show that the Sync-Transformer can achieve a comparable result with the best transformer, which is better than LAS \cite{changhaoshan2019}, RNN-T and our previous (Chunk-Flow) SA-T \cite{zhengkun2019sat}. By contrast, the Sync-Transformer can achieve online decoding with only a little degradation of the performance.  

The relationship between the decoding steps of different models is $U \leq U+M \leq T$. Most of the attention-based models, like LAS and transformer, require the least steps during inference. However, restricted by the dependencies of attention on the entire input sequence, they cannot be directly applied to online speech recognition tasks. Chunk-Flow SA-T, an RNN-free transducer model, decodes a sequence frame by frame. And it also consumes much more memory during training. However, Sync-Transformer requires less decoding steps compared with SA-T and RNN-T, which means fewer costs in memory and time.

\begin{table}[t]
	\caption{Comparisons with other models (CER \%).}
	\label{tab:window}
	\centering
	\begin{tabular}{c|cccc}
		\toprule
		\textbf{Model} & \textbf{Online} & \textbf{Steps} & \textbf{Dev} & \textbf{Test} \\
		\hline
		LAS \cite{changhaoshan2019} & No & U & - & 10.56 \\ 
		Transformer & No & U & 7.80 & 8.64 \\ 
		RNN-T \cite{zhengkun2019sat} & No & T & 10.13 & 11.82 \\ 
		SA-T \cite{zhengkun2019sat} & No & T & 8.30 & 9.30 \\ 
		Chunk-Flow SA-T \cite{zhengkun2019sat} & Yes & T & 8.58 & 9.80 \\ 
		Sync-Transformer & Yes & U+M & 7.91 & 8.91 \\
		\bottomrule
	\end{tabular}
	\vspace{-15pt}
\end{table}

\vspace{-5pt}
\section{Conclusions and Discussion}
\vspace{-5pt}

\label{sec:typestyle}

In this paper, we propose a Sync-Transformer, which combines the advantages of transformer and transducers model. In order to get rid of the dependence on the entire input state sequences, we force every node in the encoder to focus on its left context and ignore its right contexts completely during calculating self-attention. Once a fixed-length chunk of state sequence is produced by the encoder, the decoder begins to predict symbols immediately. During training, we introduce a forward-backward algorithm to sum over the probabilities of all possible alignment paths and apply a negative log-loss function to optimize Sync-Transformer. The experiments show that the Sync-Transformer can encode and decode synchronously. What's more, it outperforms our previous self-attention transducer and achieves a comparable result with the advanced transformer model. The experimental results reveal that Sync-Transformer is a very promising model for online speech recognition. However, there are some aspects to be improved in the future. For example, Sync-Transformer might raise more delete errors during inference, and recognize the wrong words with similar pronunciation. All of these prevent Sync-Transformer from being applied immediately. And it will be our next research direction.

\vspace{-5pt}
\section{Acknowledge}
\label{sec:majhead}
This work is supported by the National Key Research \& Development Plan of China (No.2018YFB1005003), the National Natural Science Foundation of China (NSFC) (No.61831022, No.61771472, No.61773379, No.61901473) and Inria-CAS Joint Research Project (No.173211KYSB2017 0061 and No.173211KYSB20190049).

\vfill\pagebreak

\bibliographystyle{IEEEbib}
\bibliography{strings,refs}

\end{document}